\documentclass[]{pasj02} 
\usepackage[switch,mathlines]{lineno} 
\usepackage{color}
\jyear{2024}
\Received{}
\Accepted{}
\usepackage[utf8]{inputenc}

\begin{document} 

\title{
{\it NuSTAR} detection of a hot stellar superflare with a temperature of 95 MK in hard X-rays}

\author{Tomohiro \textsc{Hakamata}\altaffilmark{1}\altaffilmark{*}, Hironori \textsc{Matsumoto}\altaffilmark{1,2}, Hirokazu \textsc{Odaka}\altaffilmark{1,2}, Shinsuke \textsc{Takasao}\altaffilmark{1}}%

\altaffiltext{1}{Department of Earth and Space Science, Graduate School of Science, Osaka University, 1-1 Machikaneyama, Toyonaka, Osaka 560-0043, Japan}
\altaffiltext{2}{Forefront Research Center, Graduate School of Science, Osaka University, 1-1 Machikaneyama, Toyonaka, Osaka 560-0043, Japan}
\email{hakamata@ess.sci.osaka-u.ac.jp}


\maketitle

\begin{abstract}
A search of the hard X-ray archive data of the Nuclear Spectroscopic Telescope Array ({\it NuSTAR}) found a transient source, NuSTAR J230059+5857.4, during an observation of 1E 2259+586 on 2013 April 25. A multi-wavelength analysis using X-ray, optical, and infrared data, mostly taken in its quiescent phase, was conducted to identify the origin of NuSTAR J230059+5857.4 and elucidate the phenomena associated with  the flare activity. The results indicated that NuSTAR J230059+5857.4 was a stellar flare  that occurred on a single M-dwarf, M-dwarf binary, or pre-main-sequence star. NuSTAR J230059+5857.4 exhibited the higher emission measure and higher temperature, 8.60$^{+2.15}_{-1.73} \times 10^{54}$ cm$^{-3}$ and 8.21$^{+2.71}_{-1.86}$ keV, respectively, on average than the nominal values of stellar flares reported in the past. The flare loop size estimated on the basis of the model to balance the plasma and magnetic pressures was larger than the stellar radius by a factor of several. Since based on solar flare loops, this flare loop scale is excessively large, we conjecture that the observed large emission measure is possible to be attributed to the  observation of mutually-associated multiple flares simultaneously occurring on the stellar surface, known as  sympathetic flares. Thanks to the large effective area of {\it NuSTAR} in the hard X-ray band, we are able to conduct detailed discussion about a temperature variation associated with the flare. Investigation of the temperature variation during the flare revealed that the temperature remained significantly higher than during the quiescent phase even after the count rate dropped to around 5\% of the peak. The sustained high temperature over the long duration is consistent with the idea of sympathetic flares.  We found that it is essential to use theoretical models to  evaluate loops and assess temporal changes in temperature as done in this study to determine whether there are multiple flares or not when analyzing flare observation data.
\end{abstract}


\section{Introduction}

\label{Introduction}
Stellar flares are observed on dwarf stars, which have spectral types of M, K, G, and F (Tsuboi et al. 2016). Stellar flares are also seen in pre-main-sequence stars and RS CVn systems. Pre-main-sequence stars have spectral types ranging from M to K, and RS CVn systems are close binary systems,  usually composed of G or K-type giants or subgiants and G, K, or M-type dwarfs or subgiants (Berdyugina 2005). Cool stars, including pre-main-sequence stars with M-type spectra,  have the $\alpha^{2}$ dynamo,  generating strong magnetic fields and more energetic flares than the Sun. In contrast, hot stars with G-type spectra have the $\alpha\Omega$ dynamo,  generating weaker flares (Shibata $\&$ Yokoyama 1999, 2002; Getman et al. 2022, 2023a). In RS CVn systems, tidal forces between binary stars generate strong magnetic fields and energetic flares (Berdyugina 2005).  \par

Observations of solar flares have revealed that the hard X-ray emission (above 10 keV) and soft X-ray emission (below 10 keV)  originate from the nonthermal plasma accelerated by magnetic reconnections and the thermal plasma heated by the nonthermal plasma, respectively (Veronig et al., 2005).  These facts indicate that the softer X-ray flux peak in a flare  is expected to lag behind that of harder X-rays,  as indeed observed in  solar flares (Aschwanden $\&$ Alexander 2001).  These suggest that the hard X-ray band is important for understanding  physics of stellar flares in general, such as the temporal variation of the plasma temperature in the early phase of a flare. To understand the physics of the early phase of stellar flares, it is crucial to estimate the scale of flare loops using hard X-ray data because the accelerated plasma move along the magnetic fields within the flare loops.\par

In the soft X-ray band, there have been ample detections and observations of stellar flares.   Several studies have analyzed temporal variations of the plasma temperatures of flares based on soft X-ray observations that covered the periods from the flux peaks to the decay phases of the flares (e.g., Favata et al. 2000a; Tsuboi et al. 1998). In contrast, observations in the hard X-ray band above 10 keV are very limited. In particular, none of the reported hard X-ray observations of positive detection of stellar flares have covered the entire time period of a flare from the flux peak to the decay phase and revealed a temporal variation in a temperature with high sensitivity, partly because most of them were based on   all-sky survey observations (e.g.,  Tsuboi et al. (2016), and Osten et al. (2007, 2010)), where the telescopes either have a low sensitivity or do not keep exposure for a prolonged time. For example,  Tsuboi et al. (2016) failed to measure the temperature variations associated with the flares  with a meaningful error in their MAXI observation due to the limited sensitivity of {\it MAXI}. Although Osten et al. (2007, 2010) succeeded in obtaining temperature variations of flares on II Peg and EV Lac  with {\it Swift}/BAT, the observation period did not cover the entire flare duration; in the former study, over a half of the decay phase was not observed,  and in the latter study, the observation started after the flux of the flare had already started to decline. Thus, scientists still wait for the first  high-sensitivity hard X-ray observations that cover the entire  flare period from initial rises to decay phases. \par

\begin{figure*}[htb]
 \begin{center}
  \includegraphics[keepaspectratio,scale=0.31]{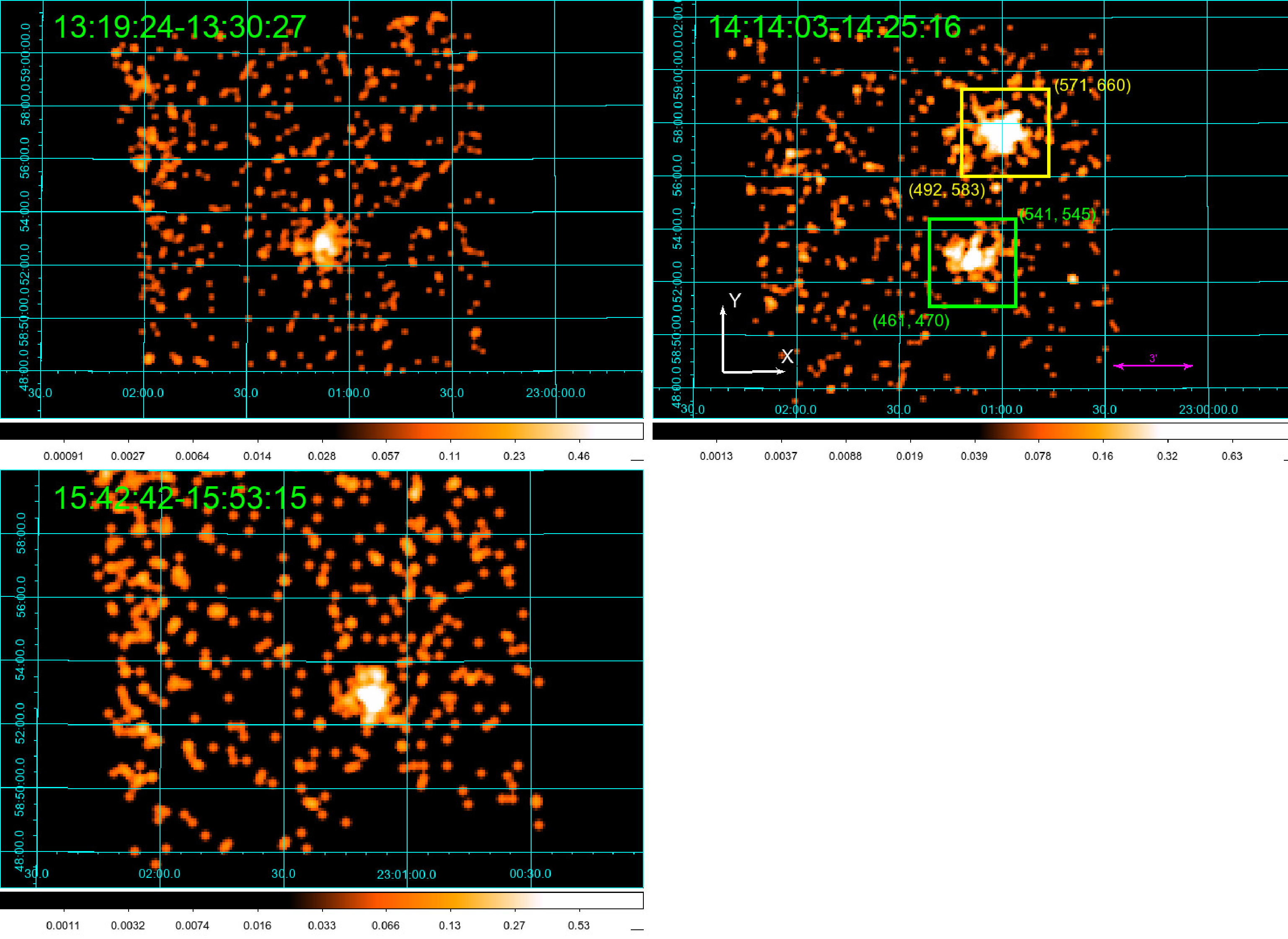}
  \end{center}
 \caption{X-ray images in the 3--79 keV band observed with {\it NuSTAR} on 2013 April 25. The  source at (RA, Dec)$_{\rm J2000.0}$ = (\timeform{23h01m08.30s}, \timeform{+58D52'44.5''}) is magnetar 1E 2259+586. The source located approximately 5$^{'}$ north of 1E 2259+586 in the upper-right panel is the transient source. In the top right panel, the regions employed for the profile fitting of 1E 2259+586 and the transient source are indicated by green and yellow squares, respectively.}
 \label{nuflare_emrge}
\end{figure*}

In this work, we searched for stellar flare candidates from the archival data of the Nuclear Spectroscopic Telescope Array ({\it NuSTAR}; Harrison et al. 2013). {\it NuSTAR} has the highest sensitivity in the hard X-ray band above 10 keV among the past and currently-operating X-ray satellites, owing to its unprecedented mirror-based X-ray imaging capability in the hard energy band extending beyond well above 10 keV (3--79 keV).  It has a large effective area, ten times larger than {\it MAXI}. The expected number of serendipitous detections of stellar flares with {\it NuSTAR} per unit time is lower than that with {\it MAXI} because {\it NuSTAR} is designed for pointing observations unlike the all-sky monitor MAXI. In fact, since its launch, {\it NuSTAR} has only detected four stellar flares in two pre-main-sequence stars and two protostars, and the detailed temperature variations of these flares have not been analyzed (Vievering et al. 2019; Getman et al. 2023b). With this search, we discovered a transient source  of which the light curve resembled a stellar flare,  located approximately 5$\arcmin$ north of the magnetar 1E 2259+586 in the observation on 2013 April 25. No source was detected at the location (see the image in Figure~\ref{nuflare_emrge}) before the flaring event, which suddenly  began at around 14:00.  The source became  faint below the {\it NuSTAR} detection threshold again approximately 600 seconds later. No prior research had described the counterpart object at the location of this transient source. This flare is hotter and brighter than the flares described in previous studies, resulting in data with many counts in the hard X-ray band thanks to the large effective area of {\it NuSTAR}. Consequently, we were able to measure the temperature variations associated with the flare and discuss these in relation to the scale of the flare loops.\par

In addition to  \textit{NuSTAR} data, we conduct  comprehensive analysis using the observation data  in both hard and soft X-rays, as well as optical and infrared bands, to  study the nature of the transient source. In this paper, we describe the observations and data reduction in the X-ray band in section \ref{Observations}, present the results of the analysis of the X-ray, optical, and infrared data in section \ref{Analysis}, and  discuss  the origin of the transient source,  the size of the flare, and the temporal temperature variation in section \ref{Discussion}.\par

\section{Observations and data reduction}
\label{Observations}

{\it NuSTAR} has two identical optical systems that incorporate two identical focal-plane modules  called the FPMA and FPMB. These focal plane modules contain CdZnTe detectors, which are sensitive to hard X-rays in  3--79 keV  (Harrison et al. 2013). {\it NuSTAR} observed the region around the transient source in four  occasions in total, three times in between 2013 April 24 and 27  and once on 2013 May 18. The transient source was detected on 2013 April 25 during the first observation period. \par

In addition to {\it NuSTAR}, the region around the transient source was observed by soft X-ray observatories, {\it Swift}, {\it Chandra}, {\it XMM-Newton}, and {\it Suzaku}. Table \ref{tab:obsdata} summarizes the X-ray observations.\par

We used Heasoft 6.28 to extract data  of {\it NuSTAR}, {\it Swift}, and {\it Suzaku}. For generating response matrix files (rmf) and ancillary response files (arf), we employed, respectively, \texttt{numkrmf} and \texttt{numkarf} for the {\it NuSTAR} data, and \texttt{xisrmfgen} and \texttt{xissimarfgen} for {\it Suzaku}.  For the {\it Swift} data, we used \texttt{xrtmkarf} for creating an arf file and simply adopted  \texttt{swxpc0to12s6$\_$20130101v014.rmf} from the CALDB database  as an rmf file. For {\it Chandra}, we used CIAO 4.13, employing \texttt{mkrmf} and \texttt{mkarf} for generating rmf and arf files, respectively. For {\it XMM-Newton}, we employed SAS 1.3 and processed the data as follows. We defined Good Time Intervals (GTIs) as time intervals when the count rates in the 10--12 keV band of the EMOS 1 detector were below 0.2 counts s$^{-1}$ to remove potential cosmic-ray-induced events. Events classified as grades 0--12 were used for analysis. We extracted photon events with the specified GTI and grade from the EMOS 1 and 2 data and created the spectra. We created rmf and arf files  with \texttt{rmfgen} and \texttt{arfgen} in SAS, respectively. For spectral analyses of any of the X-ray data, we employed XSPEC 12.11.1 in Heasoft 6.28. \par

\begin{table*}
  \tbl{Observation data used in this  work}{%
  \begin{tabular}{ccccc}
      \hline \\
      instrument & obsID & Observation start (TT)\footnotemark[$*$] & Observation end (TT)\footnotemark[$*$]& exposure [sec] \\
      \hline
      {\it NuSTAR}/FPMs A, B & 30001026002 & 2013 April 24 21:51:07 & 2013 April 25 17:16:07 & 37291.8\\
         & 30001026003 & 2013 April 25 17:16:07& 2013 April 26 00:21:07 & 15458.1\\
         & 30001026005 & 2013 April 26 22:11:07& 2013 April 27 05:36:07 & 16305.7\\
         & 30001026007 & 2013 May 16 09:31:07& 2013 May 18 06:01:07 & 88379.8\\
        {\it Swift}/XRT & 00080292002 & 2013 April 25 00:23:55& 2013 April 25 23:10:56 & 13120.8\\
        {\it Chandra}/ACIS & 725 & 2000 January 12 02:20:52 & 2000 January 12 08:15:16 & 18892.5\\
         & 6730 & 2006 April 09 09:14:03 & 2006 April 09 16:46:34 & 24783.0\\
        {\it XMM-Newton}/EPIC MOSs 1, 2 & 0038140101 & 2002 June 11 09:06:08 & 2002 June 11 23:37:03 & 51998.0\\
        {\it Suzaku}/XISs 0, 1, 3 & 404076010 & 2009 May 25 20:00:17 & 2009 May 27 15:28:14 & 89162.0\\
      \hline
    \end{tabular}}\label{tab:obsdata}
\begin{tabnote}
\footnotemark[$*$] $Suzaku$ is represented in UT. \par
\end{tabnote}
\end{table*}

\section{Analysis}
\label{Analysis}
\subsection{NuSTAR}
\subsubsection{The positional determination of the transient source}

\begin{figure}[htb]
\begin{center}
\includegraphics[keepaspectratio,scale=0.185]{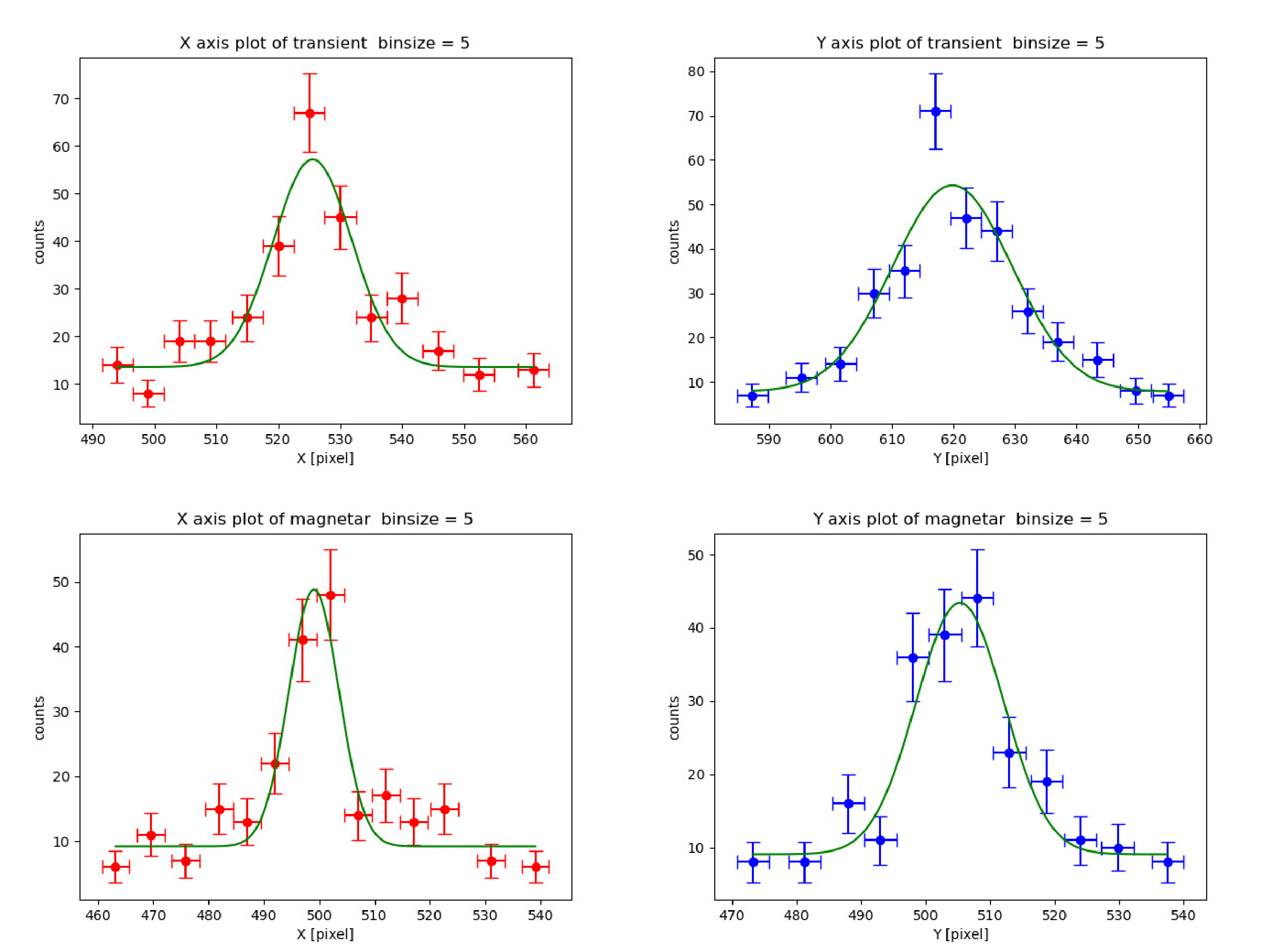}
\end{center}
\caption{Determination of the positions of the (upper two panels) transient source and (lower two panels) magnetar 1E 2259+586. The projected  profiles onto the X and Y axes are represented by red and blue crosses, respectively. The units of the horizontal and vertical axes are the pixel numbers of the $NuSTAR$ detectors and X-ray counts, respectively. Green lines show the best-fit models.}
\label{RADec_se}
\end{figure}

First of all, we determined the precise position of the transient source as follows. We first extracted a light curve from a circular region with a radius of  $1{\arcmin}\!.4$  centered at the visually determined peak at the coordinates (RA, Dec)$_{\rm J2000.0}$ = (\timeform{23h00m59.73s}, \timeform{+58D57'41.92''}). Based on the light curve, we  set the active period of the transient source as a time interval  between 14:14:03  and 14:25:16 on 2013 April 25.  We made the projected count profiles of the entire filed of view during this time interval onto the X and Y axes, which correspond to the horizontal and vertical directions in the pixel coordinates, respectively. We performed a chi-squared fit with a Gaussian function with each profile to estimate the positions of 1E 2259+586 and the transient source.  Figure \ref{RADec_se} plots the obtained profiles roughly centered at the locations of 1E 2259+586 and the transient source, overlaid with the best-fit functions. \par

To improve positional accuracy, we used data from the {\it Chandra} source catalog (CSC). We calculated the offset between the position of 1E 2259+586 recorded in the CSC and the position estimated through the fitting of the {\it NuSTAR} data. Using this offset, we corrected the position of the transient source estimated through the fitting. The {\it Chandra} positional accuracy we used for 1E 2259+586 was 0$\arcsec$.95 at a 95$\%$ confidence interval ,which is based on the most recent CALDB of {\it Chandra} as of 2022 November 11, as documented by the Chandra X-ray Center.  In consequence, the position of the transient source was determined to be (RA, Dec)$_{\rm J2000.0}$ = (\timeform{23h00m59.88s}$\pm$0$^{\rm s}$.65, \timeform{+58D57'25.86''}$\pm$6${\arcsec}\!.$14) in a 95$\%$ confidence interval. Based on the determined position, we named the transient source NuSTAR J230059+5857.4 following the IAU-approved format. \par

\begin{figure}[htb]
  \begin{center}
  \includegraphics[keepaspectratio,scale=0.12]{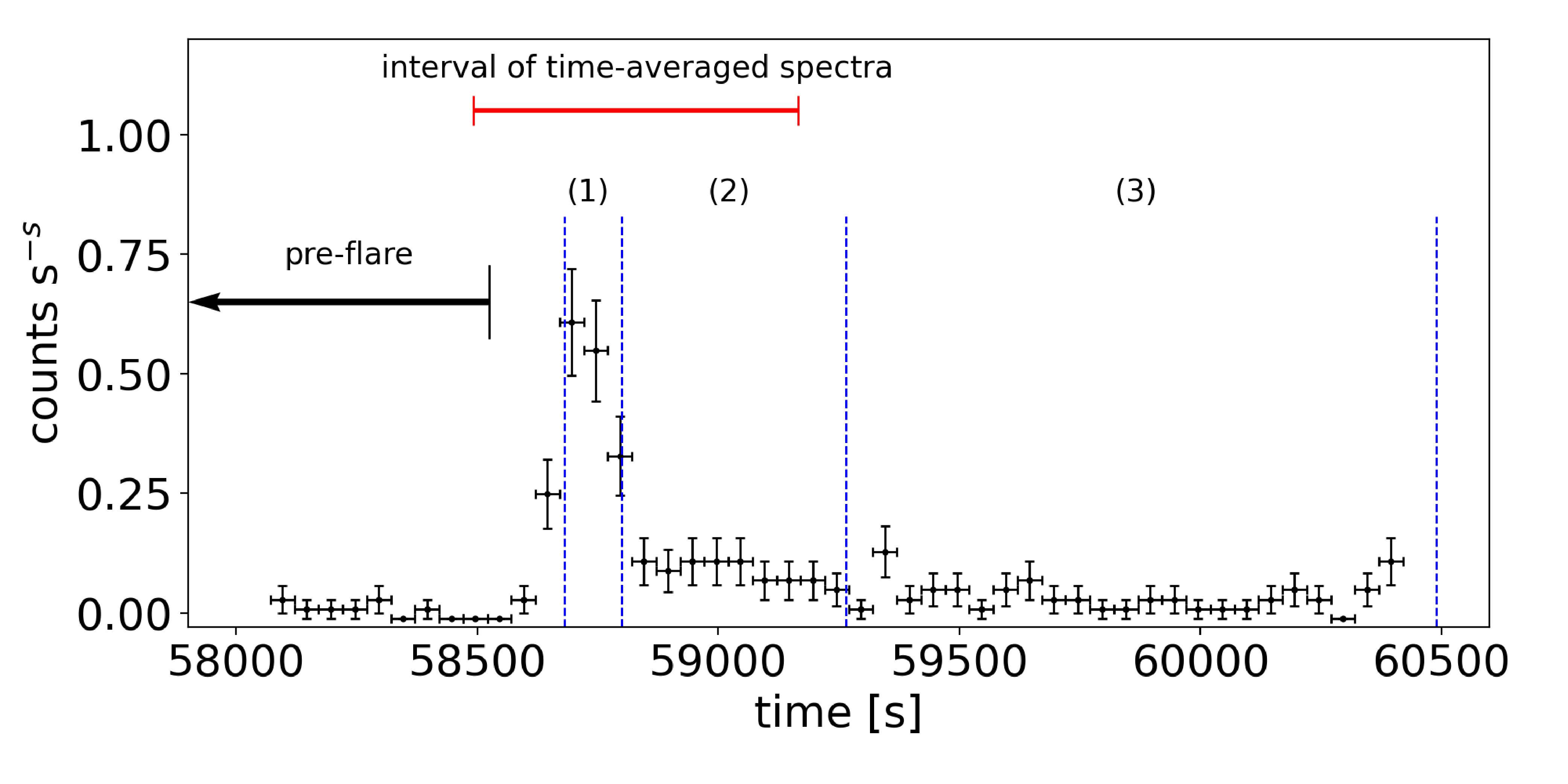}
  \end{center}
 \caption{Background-subtracted light curve extracted from the FPMA data in 3--79 keV. The black crosses  show the observed data. The horizontal red line and black arrow indicate the time intervals used to extract the time-averaged spectra and the pre-flare count rates, respectively. The blue dashed lines denote the boundaries for the time intervals of  phases (1)--(3). The origin of the time on the X-axis is at MJD = 56406.916. The count rate in the background region was $(1.19\pm0.02)\times10^{-2}$ counts s$^{-1}$.}
 \label{nuflare_lc}
\end{figure}

\subsubsection{Light-curve analysis}
We then extracted 3--79 keV light curves of NuSTAR J230059+5857.4 from the refined source region   of a circle with a 1$\arcmin$-radius centered at the  determined position.  The background region was defined as a  concentric annular region with    radii  of 2$\arcmin$--3$\arcmin$.  Figure \ref{nuflare_lc} shows the background-subtracted light curve extracted from the FPMA data. 

We measured the  e-folding time $\tau$,  regarding it as the  decay time,  of the extracted light curves in the decay phase by means of simultaneous model-fitting of the FPMA and FPMB light-curves. In the fitting, we employed C statistics (Cash 1979), in which the statistic $C$ is used instead of using $\chi^{2}$ in the chi-squared distribution,  given the low photon statistics in each bin. Therefore, for the fitting, we used the light curves without subtracting the background and employed an exponential function for NuSTAR J230059+5857.4, and a constant function to represent the background as the model. The values  of these constants were fixed at the averaged count rates in the background regions, $1.19\times10^{-2}$  and $1.03\times10^{-2}$ counts s$^{-1}$ for the FPMA and FPMB, respectively.  We allowed the normalizations of the exponential functions   for the FPMA and FPMB data to vary individually.

\begin{figure}[htb]
 \begin{center}
  \includegraphics[keepaspectratio,scale=0.37]{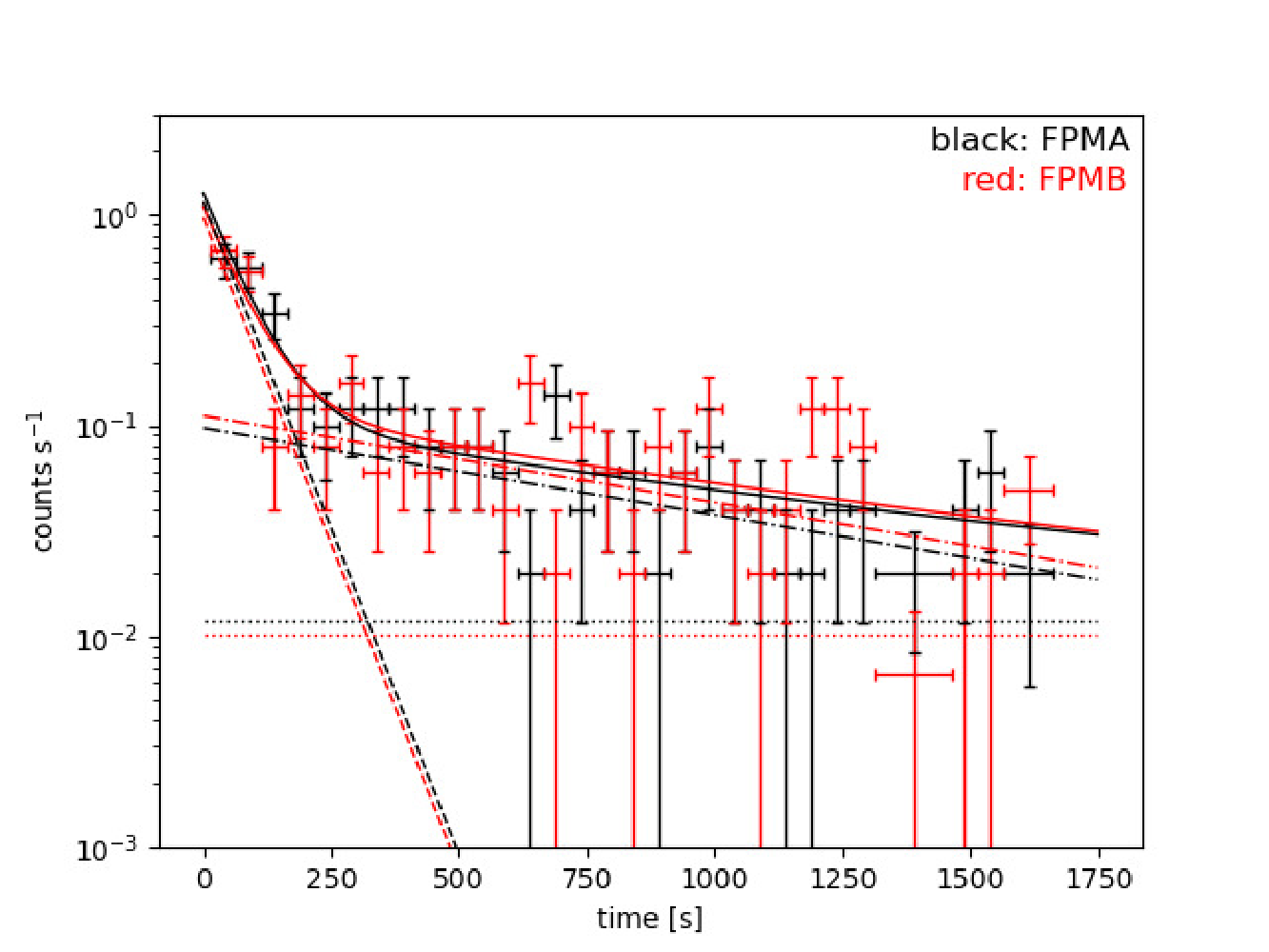}
  \end{center}
 \caption{Light curve of NuSTAR J230059+5857.4. Crosses  show the observed data,  and  solid lines  do the best-fit double-exponential model. The dashed and dot-dashed lines  show  the faster- and slower-decay components, respectively,  in the best-fit model, whereas the dotted lines,  the background. Black and red colors denote those  for the FPMA and  FPMB data, respectively.  On the horizontal axis, the origin (t=0) corresponds to the epochs of the peaks of the light curves.}
\label{decayfit}
\end{figure}

 An attempt of fitting with a single exponential function  yielded $\tau$=382.3 seconds.    However, the fit result was rejected according to the method described in Kaastra (2017) at a 5$\%$ significance level.
 The fitting with a double exponential functions     was not rejected at a 5$\%$ significance level.        Figure \ref{decayfit} and Table \ref{tab:decayfit} show the light curves with the best-fit functions and the best-fit parameters with errors, respectively. The fast and slow decay time were 70.1$\pm$27.6 and 1055.7$\pm$543.4 seconds, respectively. The fast decay was dominant  in the initial $\sim$170 seconds of the decay, from which  the slow decay rapidly became dominant.   \par

\begin{table}
  \tbl{Fitting results of NuSTAR J230059+5857.4 light curves}{
  \begin{tabular}{ccc}
      \hline \\
        & 
        \begin{tabular}{c}
        {\rm norm}\footnotemark[$*$] [{\rm counts s$^{-1}$}]\\
        \end{tabular}
        &
        \begin{tabular}{c}
        $\tau$ [sec]\\
        \end{tabular}\\
        \hline
        fast decay & 1.06$\pm$0.59 &  70.1$\pm$27.6\\
       slow decay & 0.11$\pm$0.04 & 1055.7$\pm$543.4\\
       \hline
  \end{tabular}}\label{tab:decayfit}
\begin{tabnote}
Note : Errors are at the 1 sigma confidence level. The statistic $C$ and degree of freedom are 66.48 and 54, respectively. \par
\footnotemark[$*$] Normalization of the exponential function, i.e., $A$ in  $A~\textrm{exp}(-t/\tau)$. It is averaged over the FPMA and FPMB data.\par
\end{tabnote}
\end{table}

\subsubsection{Spectral analysis}
We conducted an analysis of the time-averaged spectra  of NuSTAR J230059+5857.4. The spectra were extracted from the same region as with the light curves and  for the same time interval as for the position determination. The used time interval is shown in Figure \ref{nuflare_lc}. Each spectrum exhibited an emission line at around 6.7 keV, which is often observed in optically-thin thermal plasma. The spectra with the FPMA and FPMB were simultaneously fitted  with the {\it Bremss} model in \textsc{XSPEC}  for the continuum and a Gaussian function for the emission line,  modified with the {\it TBabs} model to correct for absorption caused by interstellar matter.
Given that {\it NuSTAR} is sensitive in the energy band of 3--79 keV, in which  the effect of absorption is   limited, we fixed the column density to the value 5.9 $\times$10$^{20}$~cm$^{-2}$ determined from the {\it Chandra} observation data (see section \ref{Other_X-ray_observatories}).
The redshift was fixed at zero because observations of {\it GAIA}  determined the distance to NuSTAR J230059+5857.4 to be 281 pc, indicating that it is located  inside the Milky Way galaxy (see section \ref{Optical and infrared observations}). Parameters were linked between the FPMA and  FPMB data with a normalization scaling factor between them because there is the maximum systematic calibration error of about 5\% in those normalizations (e.g. Harrison et al. 2013).  The factor of FPMA was fixed at one, while that of FPMB was left free. A difference between constants was 4.2 \% as a result. Consequently, the center energy of the emission line in the spectra was determined to be 6.69$^{+0.17}_{-0.13}$ keV within a 90$\%$ confidence interval. It means that the line is the He-like Fe-K$\alpha$ line. \par

\begin{figure}[htb]
  \includegraphics[keepaspectratio,scale=0.3, angle=270]{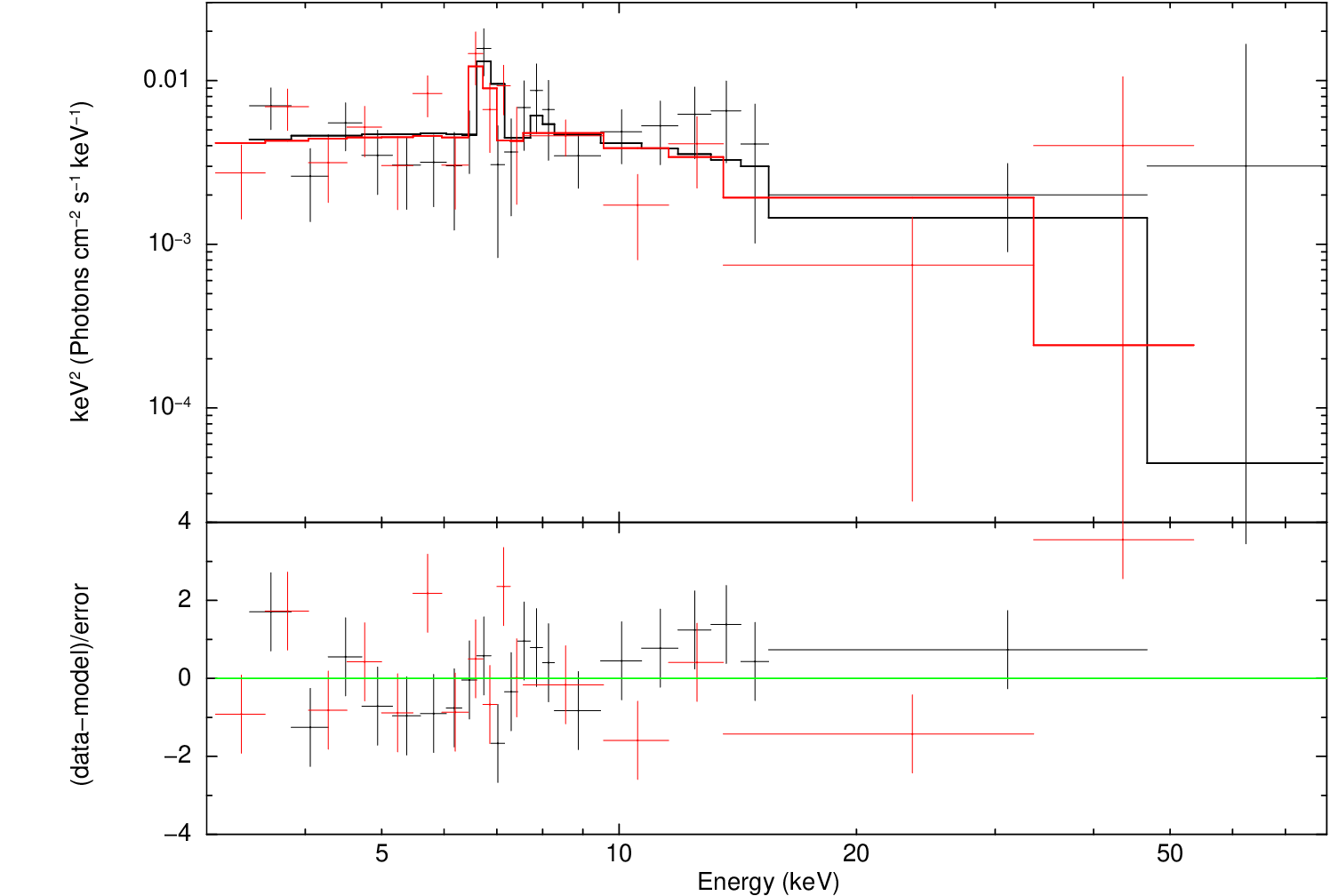}
 \caption{X-ray spectra of NuSTAR J230059+5857.4. Black and red colors denote the data and fitting results  for the FPMA  and FPMB, respectively.}
 \label{nucosp}   
\end{figure}

Since the He-like Fe-K$\alpha$ line was observed in the spectra, the X-rays from NuSTAR J230059+5857.4 were likely to be originated from optically-thin hot plasma. Therefore, we fitted the spectra with the {\it APEC} model in XSPEC. Similar to the fitting with the {\it Bremss} model, the parameters were linked except the {\it APEC} normalization that represents an emission measure ($EM$). We quantified the unabsorbed flux in the 0.5--10.0 keV band, using the $cflux$ model, which is the convolution model to calculate the flux of components in the model in XSPEC,  and derived the total radiative energy in the 0.5--10.0 keV band, $E_{tot}$,  combining the obtained flux with the distance obtained from {\it GAIA}. Figure \ref{nucosp} shows the spectra  overlaid with the best-fit model, and  Table \ref{tab:integspecfit} summarizes the fitting result. The temperature and emission measure are higher than these of stellar flares reported in previous studies, in which the best-fit temperatures and emission measures are typically 1-6 keV and 10$^{52-54}$ cm$^{-3}$, respectively. (see section \ref{Scale of the flare}).

\begin{table}
  \tbl{Fitting results of the time-averaged spectra\footnotemark[$*$]  of NuSTAR J230059+5857.4}{
  \begin{tabular}{cccc}
      \hline \\
       Model & Parameter & Unit &  Values\\
      \hline
       APEC & {\it kT} & keV & 8.21$^{+2.71}_{-1.86}$\\
        & Abundance &  & 0.83$^{+1.00}_{-0.52}$\\
        & {\it EM} & $10^{53}~{\rm cm}^{-3}$ & 85.97$^{+21.52}_{-17.34}$\\
        cflux & flux & $10^{-13}~{\rm erg}~{\rm s}^{-1}~{\rm cm}^{-2}$ & 133.06$^{+15.49}_{-17.91}$\\
        & {\it $E_{tot}$} & $10^{34}~{\rm erg}$ & 7.9$^{+0.92}_{-1.06}$\\
        & {\it C (d.o.f.)}\footnotemark[$\dagger$]  & & 37.72 (34) \\
       \hline
  \end{tabular}}\label{tab:integspecfit}
\begin{tabnote}
Note: Errors are at the 90\% confidence level. The column density is fixed to 5.9$^{+3.1}_{-2.4}$ $\times$10$^{20}$~cm$^{-2}$. \par
\footnotemark[$*$] The time interval of the time-averaged spectra is MJD 56407.5931--56407.6009. \par
\footnotemark[$\dagger$] $C$ and $d.o.f.$ are the statistic in C statistics and the degree of freedom, respectively. \par
\end{tabnote}
\end{table}

\begin{figure}[htb]
  \includegraphics[keepaspectratio,scale=0.3, angle=270]{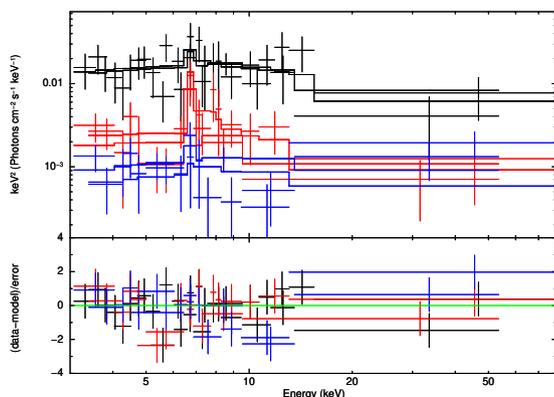}
 \caption{Time resolved X-ray spectra of NuSTAR J230059+5857.4. The data points and lines show the observed spectra and best-fit models, respectively. The spectra in phases (1), (2), and (3) (see Fig.~\ref{nuflare_lc}) are shown in black, red, and blue, respectively.}\label{numpha_spec}
 \end{figure}

To  search for temporal variations in temperature, we extracted spectra from  three  separate time intervals as indicated in phases (1)--(3) in Figure \ref{nuflare_lc}. The time intervals (1)-(3) correspond to the periods of MJD 56407.5952-56407.5966, 56407.5966-56407.6019, and 56407.6019-56407.6161, respectively. We fitted these spectra with the {\it APEC} model in the same manner  as with the time-averaged spectra.  Figure~\ref{numpha_spec} shows the spectra and fitting results, and Table~\ref{tab:coplres} summarizes the best-fit parameters. Table~\ref{tab:coplres} also includes the background-subtracted count-rate of the spectrum   in the pre-flare  time interval between 12:47:41 and 14:14:36 on 2013 April 25, the data of which have too poor statistics to constrain the spectral parameters. The count rate remained higher in all phases (1)--(3) than in the pre-flare phase. The temperature during the flare was around 10 keV, and higher than the quiescent phase temperature, $\sim$1 keV (see section \ref{Other_X-ray_observatories}) even in the darkest phase (phase (3)).  \par

\begin{table*}
  \tbl{Fitting results of the time-divided spectra of the {\it NuSTAR} observations}{
   \begin{tabular}{cccccc}
      \hline \\
       Phase & {\it kT} [keV] & Abundance & {\it EM} [10$^{53}$~cm$^{-3}$]& {\it C (d.o.f.)}\footnotemark[$\dagger$] & count rate [10$^{-2}$ cts s$^{-1}$]\footnotemark[$*$]\\
      \hline
       pre-flare & - & - & - & - & $\leq$ 0.72\\
       (1) & 10.85$^{+6.66}_{-3.13}$ & $\geq$ 2.07 & 282.15$^{+82.61}_{-69.17}$ & 20.95 (26) &61.57$\pm$17.45 \\
       (2) &  8.11$^{+5.46}_{-2.90}$ & $\geq$ 0.78 & 32.16$^{+21.18}_{-10.02}$ & 13.67 (16) & 9.75$\pm$3.70\\
       (3)  & 13.99$^{+48.09}_{-8.74}$ &  $\geq$ 0 & 15.73$^{+9.83}_{-6.19}$ & 18.54 (14) & 3.93$\pm$1.56\\
       \hline
  \end{tabular}}
\label{tab:coplres}
\begin{tabnote}
The parameter {\rm nH} of {\it TBabs}  is fixed to the value extracted from {\it Chandra} observations, 5.9 $\times 10^{20}$ {\rm cm$^{-2}$},  and redshift was fixed to 0. The errors are at the 90\% confidence level. \\
\footnotemark[$*$] Count rate in the 3--79 keV band. The parameters are averaged over the FPMA and FPMB data. \par
\footnotemark[$\dagger$] $C$ and $d.o.f.$ are the statistic in C statistics and the degree of freedom, respectively. \par
\end{tabnote}
\end{table*}

We analyzed data from the three additional {\it NuSTAR} observations listed in Table \ref{tab:obsdata}. We extracted the histograms of count rates from these data to see the amount of  flux increase  in the transient  activity. No source was detected during these observations, allowing us to determine only the upper limits of the source flux. We converted extracted count rates to flux using the {\it APEC} model, keeping most parameters fixed except for the flux of the $cflux$ model in XSPEC. We set the temperature and abundance to the average of the measured values in the two {\it Chandra} observations (as detailed in section \ref{Other_X-ray_observatories}). The column density and redshift were fixed  at the values used in the analysis of NuSTAR J230059+5857.4. During the observation  of obsID 30001026007, bursts occurred, which were observed across the entire detector surface twice; they appeared to be associated with M-class solar flares  as listed in the Hinode Flare Catalogue (Watanabe et al. 2012). Therefore, we  excluded these two  time intervals  in the analysis. The results of this analysis are presented in Table \ref{tab:nuquitfit}.  In consequence, the flux of NuSTAR J230059+5857.4 shown in Table \ref{tab:integspecfit}  was found to be at least 20 times higher than that during the other $NuSTAR$ observations with no detection of the transient source.\par

\begin{table}
  \tbl{Upper limits for the flux of the {\it NuSTAR} observations with no transient source detection}{
   \begin{tabular}{cc}
      \hline \\
       ObsID & flux [10$^{-13}$~erg~s$^{-1}$~cm$^{-2}$]\footnotemark[$*$]\\
       \hline
       30001026003 & $\leq$ 3.39 \\
       30001026005 & $\leq$ 6.31 \\
       30001026007 & $\leq$ 2.09 \\
       \hline
  \end{tabular}}\label{tab:nuquitfit}
\begin{tabnote}
The upper limits are at the 90\% confidence level. In the model-fitting,  the temperature and abundance  are fixed to 1.00~keV and 5.5$\times 10^{-2}$, respectively. \par
\footnotemark[$*$] The flux in the 0.5--10~keV band. \par
\end{tabnote} 
\end{table}

\subsection{Other X-ray observatories}
\label{Other_X-ray_observatories}

\begin{figure*}[htb]
  \includegraphics[keepaspectratio,scale=0.21]{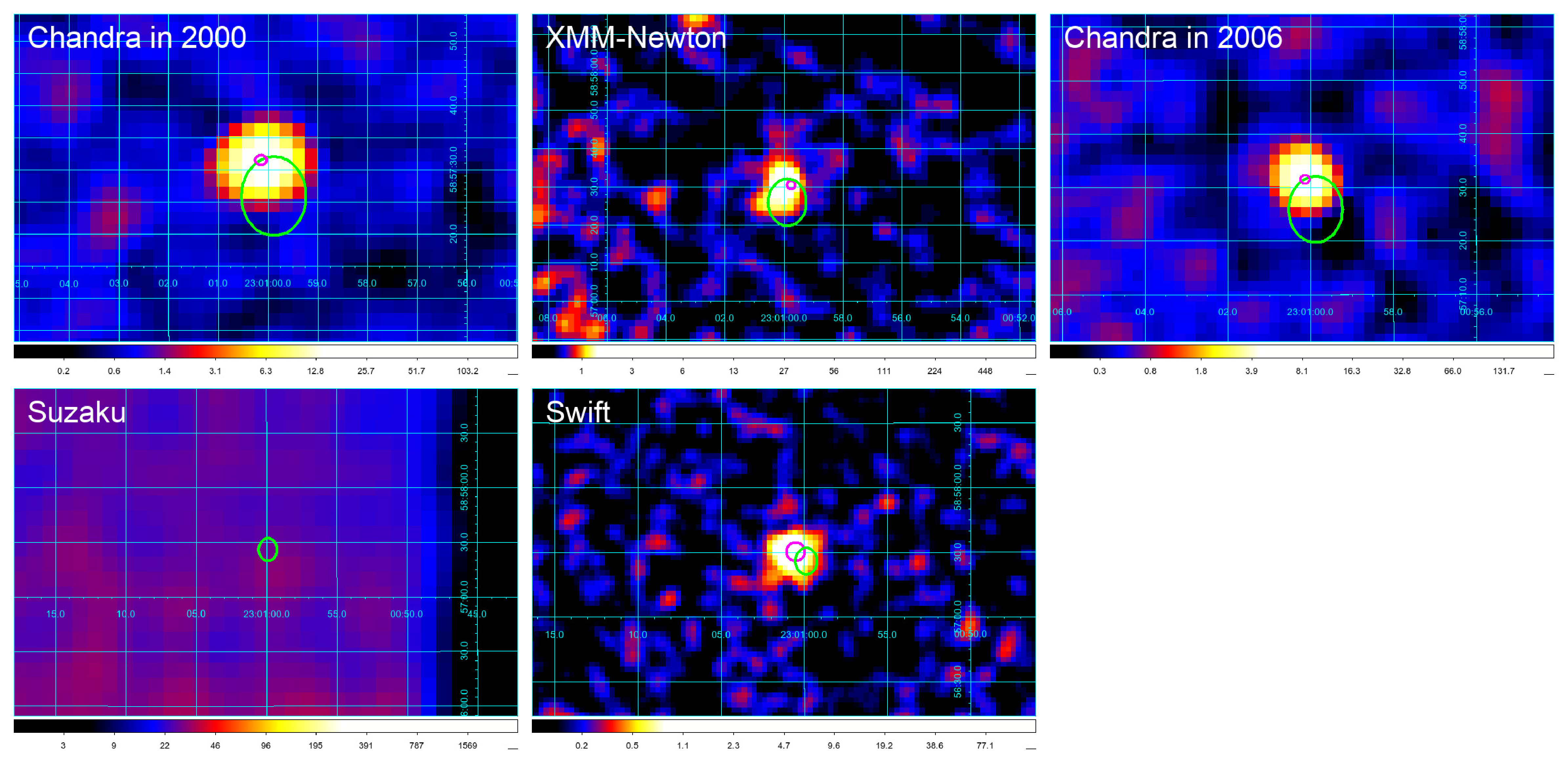}
 \caption{X-ray images from {\it Swift}, {\it Chandra}, {\it XMM-Newton}, and {\it Suzaku} data. The green and magenta ellipses  indicate the position of NuSTAR J230059+5857.4 determined using {\it NuSTAR} data and the positions of the object in each catalog, respectively. The sizes of the ellipses correspond to the positional accuracies within a 95$\%$ confidence interval.} 
\label{othXsao}
\end{figure*}

We analyzed data obtained from other X-ray observatories of {\it Swift}, {\it Chandra}, {\it XMM-Newton}, and {\it Suzaku} as listed in Table \ref{tab:obsdata}. From these data, we extracted images as shown in Figure \ref{othXsao}, and identified a source at the position of NuSTAR J230059+5857.4 with {\it Swift}/XRT, {\it Chandra}/ACIS, and {\it XMM-Newton}/EMOSs 1 and 2, but not with {\it Suzaku}/XISs (the source was detected with neither of XISs 0 and 3 and  was outside the  field of view of XIS 1 in the {\it Suzaku} observation). We cross-referenced catalogs, Swift XRT Point Source (2SXPS) Catalog, CSC, and the 2XMM catalog. These catalogs identified celestial objects known as 2SXPS J230100.4+585730, 2CXO J230100.1+585731, and 2XMM J230059.9+585730, respectively,   at the position of NuSTAR J230059+5857.4 at a 95$\%$ confidence interval. It is highly likely that the sources  cataloged by {\it Swift}, {\it Chandra}, and {\it XMM-Newton}  are associated with NuSTAR J230059+5857.4.  \par

We extracted light curves and spectra from {\it Swift}, {\it Chandra}, {\it XMM-Newton}, and {\it Suzaku}. For the observations  with {\it Swift}, {\it Chandra}, and {\it XMM-Newton}, we defined the source regions as the circles with radii of 30$\arcsec$, 6$\arcsec$, and 13$\arcsec$, respectively  centered  at the positions recorded in their respective catalogs. The background regions were defined as the concentric annuli  of the source region with inner and outer radii of 60--120$\arcsec$, 12--42$\arcsec$, and 25--75$\arcsec$, respectively.  As for the {\it Suzaku} observations, we used the circle as the source region centered at the same  position as for NuSTAR J230059+5857.4 observed by {\it NuSTAR}  with a radius of  30$\arcsec$. For the {\it Suzaku} background region, we used the circular region with the same radius as with the source region, located at an equi-distance from the persistent source 1E 2259+586, to minimize the influence of the  potential contamination from 1E 2259+586, which may be significant with \textit{Suzaku}, given  its relatively large angular resolution.  \par

\begin{figure*}[htb]
  \includegraphics[keepaspectratio,scale=0.243]{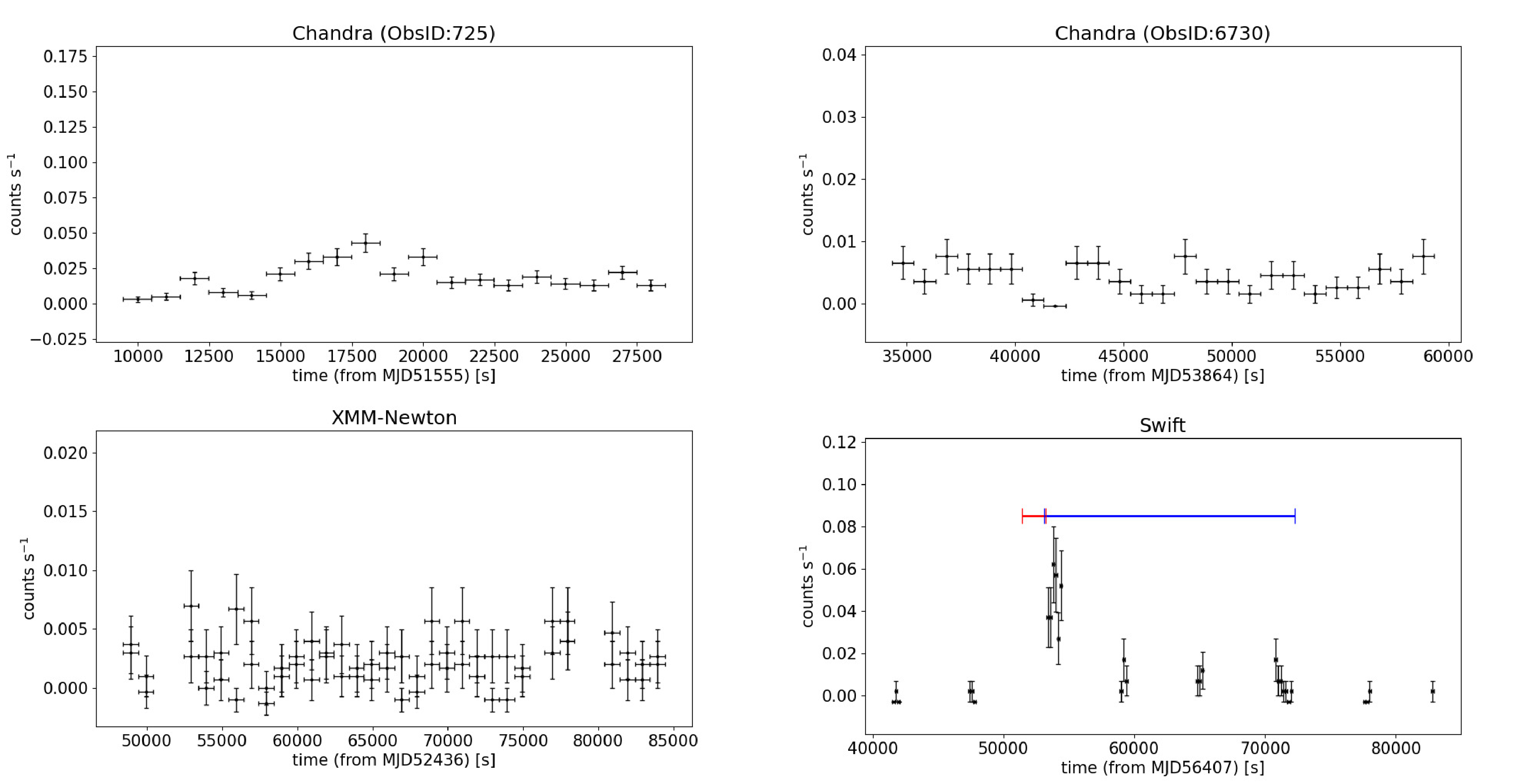}
 \caption{Background-subtracted light curves of {\it Chandra} in 2000 (upper left  panel), in 2006 (upper right  panel), {\it XMM-Newton} (lower left  panel) and {\it Swift} (lower right  panel). The energy bands  of the light curves for {\it Swift}, {\it Chandra}, and {\it XMM-Newton}  are 0.2--10.0 keV, 0.2--7.0 keV, and 0.2--12.0 keV, respectively. The  origin of time (time = 0) corresponds to MJD = 56406.916 for {\it Swift}, 51555.000 for {\it Chandra} in 2000, 53864.000 for {\it Chandra} in 2006, and 52436.000 for {\it XMM-Newton}. Error bars indicate 1 sigma errors. In the {\it Swift} light curve, the red and blue guide lines  indicate the time intervals used for the spectral analysis of {\it NuSTAR} and {\it Swift}, respectively.}\label{otherXlc}
\end{figure*}

Figure \ref{otherXlc} shows the background-subtracted light curves of the detected sources.  The time-averaged background count rates were  (3.00$\pm$0.14)$\times10^{-3}$ counts s$^{-1}$ for {\it Swift}, (1.06$\pm$0.04)$\times10^{-3}$ counts s$^{-1}$ for {\it Chandra} in 2000, (4.28$\pm$0.19)$\times10^{-4}$ counts s$^{-1}$ for {\it Chandra} in 2006, and (2.17$\pm$0.04)$\times10^{-3}$ counts s$^{-1}$ for {\it XMM-Newton} (the average of EMOSs 1 and 2).    No rapid brightening of the source similar to  that observed by {\it NuSTAR}  is found in the {\it Chandra} or {\it XMM-Newton} observations.   Notably, {\it Swift} observations  covered a period of the decay phase immediately after the {\it NuSTAR} observation, when the source was significantly brighter than in its quiescent state.\par

We  proceeded to spectral analysis.  We  determined the GTI as follows; for the {\it Chandra} in 2000 and 2006, and {\it Suzaku} data, we used the entire observation period, MJD~=~51555.116--51555.324, 53864.403--53864.681, and 54976.834--54978.644, for {\it XMM-Newton}, we adopted the time interval outlined in section \ref{Observations} within MJD = 52436.566--52436.971, and for {\it Swift}, we used the time intervals MJD~=~56407.615--56407.837 as shown in Figure \ref{otherXlc}. 

\begin{figure}[htb]
  \includegraphics[keepaspectratio,scale=0.33, angle=270]{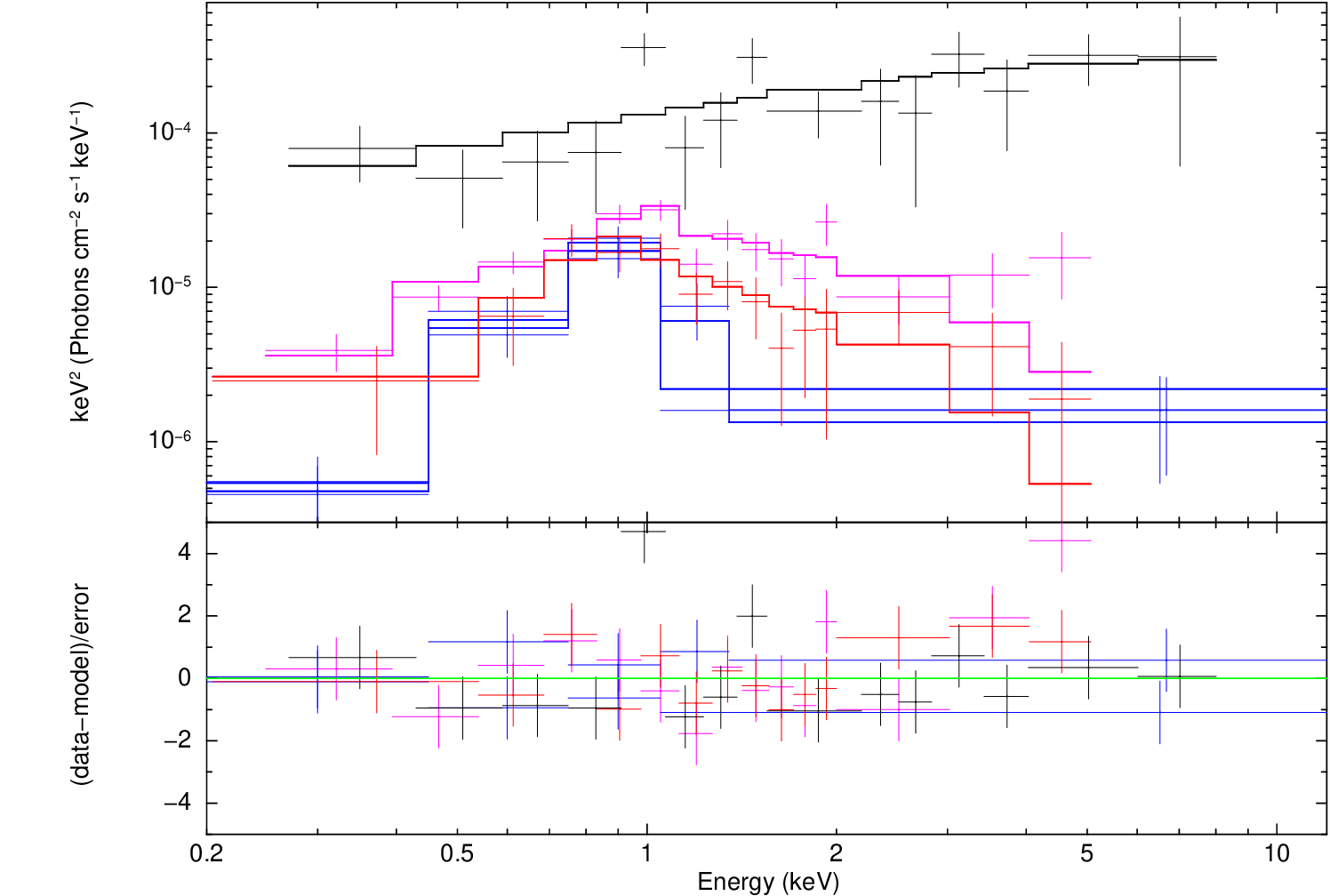}
 \caption{The spectra obtained from {\it Swift} (black), {\it Chandra} in 2000 (magenta), {\it Chandra} in 2006 (red) and {\it XMM-Newton} (blue). The crosses and lines  show the observed data and best-fit models, respectively.}\label{swspec}
\end{figure}

In our analysis, we used  energy bands  of 0.2--10.0 keV, 0.2--7.0 keV, 0.2--12.0 keV, and 0.4--12.0 keV for {\it Swift}, {\it Chandra}, {\it XMM-Newton}, and {\it Suzaku}, respectively.  We fitted the spectra with  an optically-thin thermal plasma model similar to that used in the spectral analysis of the {\it NuSTAR} data.  In the spectral fitting, the redshift were fixed at zero.  As for the column density, we allowed it to vary  for the {\it Swift} and {\it Chandra} data,  but fixed it for the {\it XMM-Newton} data   to the best-fit value  from the model-fitting of the {\it Chandra} data in 2000  because the data statistics are too poor to give a significant constraint on the fitted parameters otherwise.   The parameters  for the EMOSs 1 and 2 data were linked except for the normalization of the {\it APEC} model.  With the {\it Suzaku} data, where the source was not detected, we estimated only  the upper limit of the flux with the column density of   the best-fit value  from the {\it Chandra} data in 2000 and the temperature and abundance of  those of the averages of the best-fit values obtained from the {\it Chandra} data in 2000 and 2006.  Figure \ref{swspec} shows the spectra with the best-fit models, whereas  Table \ref{tab:otherXpar} tabulates the fitting results. The flux and temperature during the {\it Swift} observations were higher than those during the other observations. No significant difference was found among the parameters observed with {\it Chandra}, {\it XMM-Newton}.  The upper limit for the flux from the {\it Suzaku} data was not significantly different from that of {\it Chandra} and {\it XMM-Newton} data.  \par

\begin{table*}
\tbl{Fitting results of the {\it Swift}, {\it Chandra}, {\it XMM-Newton}, {\it Suzaku} data}{
\begin{tabular}{ccccccc}
\hline \\
Instrument & ${\it N}_{H}$ [10$^{20}$~cm$^{-2}$] & $kT$ [keV] & Abundance & $EM$ [10$^{53}$~cm$^{-3}$] & flux [10$^{-13}$~erg~s$^{-1}$~cm$^{-2}$]\footnotemark[$*$]& $C (d.o.f.)$\footnotemark[$\dagger$]\\
\hline
{\it Swift} & $\leq$~7.3 & $\geq$~3.67 & $\geq$~0 & 5.60$^{+4.85}_{-3.29}$ & 10.13$^{+3.04}_{-2.73}$ & 22.08 (11)\\
{\it Chandra} 2000 & 5.9$^{+3.1}_{-2.4}$ & 1.16$^{+0.17}_{-0.16}$ & 0.07$^{+0.07}_{-0.04}$ & 1.03$^{+0.32}_{-0.22}$ & 0.65$^{+0.10}_{-0.09}$ & 20.91 (11)\\
{\it Chandra} 2006 & $\leq$~27.28 & 0.85$^{+0.33}_{-0.26}$ & 0.04$^{+0.10}_{-0.04}$ & 0.73$^{+0.10}_{-0.34}$ & 0.35$^{+0.33}_{-0.09}$ & 8.86 (10)\\
{\it XMM-Newton}& 5.9 & 0.63$^{+0.16}_{-0.15}$ & 0.25$^{+0.94}_{-0.16}$ & 0.23$^{+0.13}_{-0.12}$ & 0.23$^{+0.04}_{-0.04}$ & 2.6 (5) \\
{\it Suzaku} & 5.9 & 1.00 & 0.06  & - & $\leq$~0.35 & 12.11 (6) \\
\hline
\end{tabular}}\label{tab:otherXpar}
\begin{tabnote}
Errors are at the 90\% confidence level. \par
\footnotemark[$*$] Flux in the 0.5--10.0 keV band. The fluxes were obtained  with the $cflux$ model, and the other parameters were obtained through fitting in which the $cflux$ model was not included.\par
\footnotemark[$\dagger$] $C$ and $d.o.f.$ are the statistic in C statistics and the degree of freedom, respectively. \par
\end{tabnote}
\end{table*}

\begin{figure}[htb]
  \includegraphics[keepaspectratio,scale=0.17]{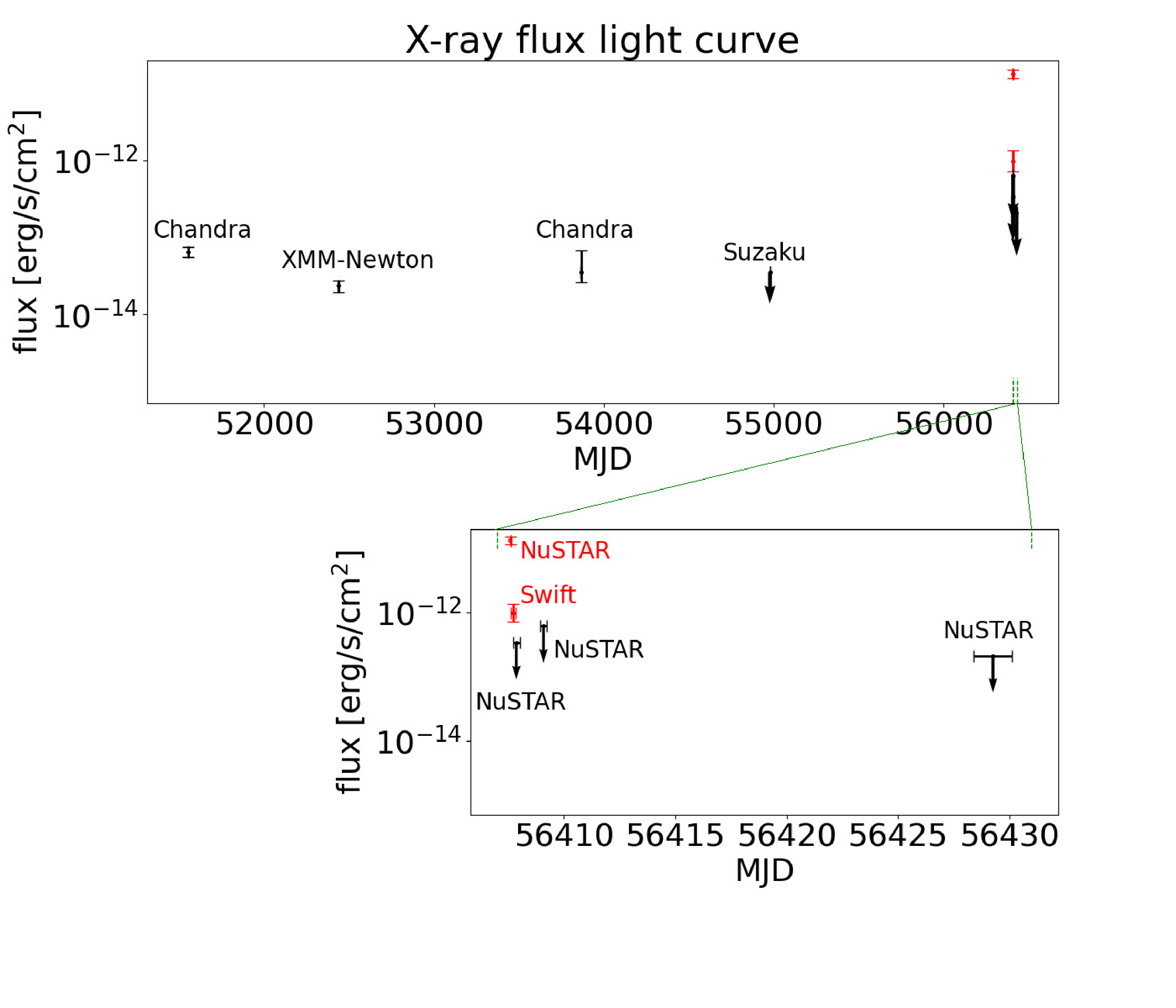}
 \caption{Long-term variation of the X-ray flux in the 0.5--10.0 keV band. The lower panel  shows a magnified view of the upper panel, encompassing the period of the transient activity. The red crosses denote the observations  with positive detection of the transient source conducted  with {\it NuSTAR} and {\it Swift},  whereas  black crosses and arrows  do the observations of quiescent emissions. The vertical and horizontal error bars indicate the flux errors at  the 90$\%$ confidence interval and the  observation periods, respectively.}\label{Xflux}
\end{figure}

Figure~\ref{Xflux} shows the obtained long-term variation of the X-ray flux in the 0.5--10.0 keV band.  With reference to the fluxes  observed  with {\it NuSTAR} and {\it Swift} at MJD = 56407, the fluxes  in the other observations were consistently lower than them and remained constant over several years.  This  implies that {\it Chandra}, {\it XMM-Newton}, and {\it Suzaku} observed quiescent-phase emissions. The quiescent level of the X-ray flux in the 0.5--10.0 keV band was approximately 4 $\times$10$^{-14}$~erg~s$^{-1}$~cm$^{-2}$, which means that the flux increased by a factor of up to 320   during the transient activity of NuSTAR J230059+5857.4. \par

\subsection{Optical and infrared observations}
\label{Optical and infrared observations}

To determine the origin of NuSTAR J230059+5857.4, we investigated data observed in the optical and infrared bands.  {\it GAIA}, {\it SDSS}, {\it 2MASS}, and {\it WISE} observed regions near NuSTAR J230059+5857.4. In the catalogs  of {\it GAIA} Data Release 3, SDSS DR10, 2MASS All-Sky Point Source Catalog, and AllWISE Data Release, celestial objects were found at coordinates (RA, Dec)$_{\rm J2000.0}$ = (\timeform{23h01m00.10s}, \timeform{+58D57'31.18''}), (\timeform{23h01m00.08s}, \timeform{+58D57'31.20''}), (\timeform{23h01m00.07s}, \timeform{+58D57'31.11''}), and (\timeform{23h01m00.09s}, \timeform{+58D57'31.23''}), respectively. All of these positions were consistent with the location of NuSTAR J230059+5857.4 at a 95 $\%$ confidence interval. Therefore, the objects observed in these missions are likely to be the counterpart of the source of NuSTAR J230059+5857.4.\par

We utilized archival data from {\it GAIA}  to measure a distance to the counterpart of the source of NuSTAR J230059+5857.4. {\it GAIA} has a high parallax accuracy ranging from 25 to 347 $\mu$as (Bailer-Jones et al. 2013), enabling direct distance measurement from observed parallaxes. {\it GAIA} observed the parallax of the counterpart and determined the distance to be 281 pc.\par

We obtained a spectral energy distribution (SED) of the infrared and optical counterpart.   Since the observed flux densities  are affected from interstellar absorption,  we applied an absorption correction to  estimate the intrinsic flux densities using the absorption  value computed  from the column density obtained from the {\it Chandra} data, where the conversion was based on the extinctions and the conversion formula described in Yuan, Liu, and Xiang (2013) and Bohlin, Savage, and Drake (1978), respectively. Additionally, we extracted light curves from single-exposure {\it WISE} images in the {\it W1} (3.4 {\rm $\mu$m}), {\it W2} (4.6 {\rm $\mu$m}), and {\it W3} (12 {\rm $\mu$m}) bands to assess errors arising from the intrinsic long-term variability of the celestial object. An error due to a variability for each band was defined as the standard deviation-to-mean ratio of flux densities in the light curve. We extracted light curves from 292, 245, and 27 images and obtained ratios of 6.2 $\%$ 8.5 $\%$, and 64.1 $\%$ for the {\it W1}, {\it W2}, and {\it W3} bands, respectively. Note that no time variability was considered for the {\it SDSS} and {\it 2MASS} data because  no light curves  of the source were available in these missions, which might result in an underestimation of the errors  of the flux densities. \par

\begin{figure*}[htb]

  \begin{minipage}{0.5\hsize}
  \includegraphics[keepaspectratio,scale=0.465, angle=0]{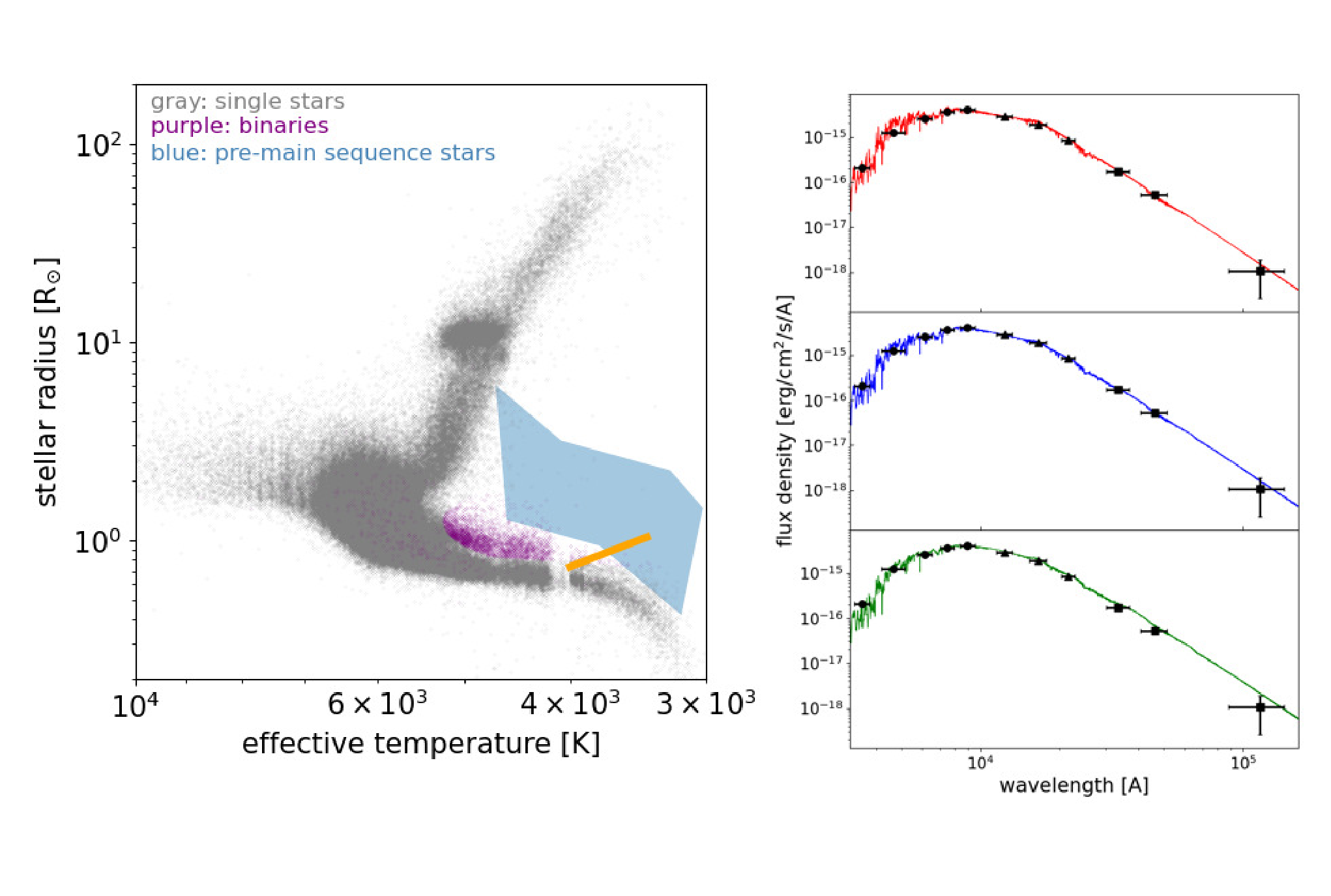}
 \end{minipage}
  \caption{Analysis results of the optical and infrared data. In the left panel,  gray and purple dots  show single and binary Kepler stars  cataloged in {\it GAIA} DR2 (Mathur et al. 2017; Berger et al. 2018), respectively. The blue-shaded  region  indicates the  region of pre-main sequence stars (stars younger than 10$^{7}$ years) with temperatures and radii of approximately 3000--4500 K and 0.3--10 R$_{*}$ on the basis of the Hayashi track (Feigelson $\&$ Montmerle 1999). The orange line represents the temperature and radius range that  reproduces the SED reasonably well. The right panels show the SEDs and models. Circular, triangular, and square  markers  denote the observed data from {\it SDSS}, {\it 2MASS}, and {\it WISE}, respectively. Vertical and horizontal error bars represent 1 sigma errors and bandwidths of photometries, respectively. The red, blue, and green lines  show the models of a single M-dwarf, an M-dwarf binary, and a pre-main-sequence star, respectively. }
 \label{fludfit}
\end{figure*}

We evaluated the SED  with a stellar radiation model, specifically the BT-NextGen model (Allard et al. 2011), to  estimate the temperature and radius of the star. We searched for parameters  that could represent the SED through visual inspection of the SED and the model; we note that a statistical evaluation of the model fitting makes little sense in this case  because the precise error estimation on some data points in the SED  has not been conducted, hence our choice of visual inspection.      Our  inspection revealed that models with temperatures  between 3400  and 4000 K and radii between 0.73 and 1.04 R$_{\odot}$  reproduce the SED reasonably well. The orange line in the left panel of  Figure \ref{fludfit} shows the temperature and radius range that effectively  reproduces the SED,  overlaid on the data of known stars.

 The three sets of the temperatures and radii shown in Table \ref{tab:opIRfitres} were taken from the orange line as examples; those parameter sets correspond to a single M-dwarf star (4000~K and 0.73~R$_{\odot}$), an M-dwarf binary (3800~K and 0.79~R$_{\odot}$),  and a pre-main-sequence star (3500~K and 0.96~R$_{\odot}$). We found that  the BT-NextGen models with these parameters can  reproduce  the SED reasonably well (see Figure~\ref{fludfit}). \par

\begin{table}
  \tbl{ Plausible models  that reproduce the optical and infrared SED}{
  \begin{tabular}{ccccc}
      \hline \\
          temperature [K] & radius [R$_{\solar}$] & star\footnotemark[$*$]\\
      \hline
      4000 & 0.73 & single  M-dwarf\\
      3800 & 0.79\footnotemark[$\dagger$] & M-dwarf binary \\
       3500 & 0.96 & pre-main-sequence star\\
       \hline
  \end{tabular}}\label{tab:opIRfitres}
\begin{tabnote}
\footnotemark[$*$] Stellar type  corresponding to  the temperature and radius.\\
\footnotemark[$\dagger$] The radius derived from the combined luminosity. If the binary system includes two stars, the radius of each star is 0.56 $R_{\solar}$.\\
\end{tabnote}
\end{table}

\section{Discussion}
\label{Discussion}
\subsection{Classification of NuSTAR J230059+5857.4}
\label{Classification of the transient source}
We consider the classification of NuSTAR J230059+5857.4. Specifically we consider, as its candidate, a stellar flare, an X-ray burst, a tidal disruption event, a supernova, and an X-ray afterglow. Here is a brief summary of these events.  The stellar flare is typically characterized by timescales of  $10^0$--10$^2$ kiloseconds in X-ray bands (e.g., G\"{u}del et al. 2004; Tsuboi et al. 2016) and occasionally as short as several hundred seconds (e.g., Osten et al. 2005). A stellar flare exhibits a spectrum that is approximated by an optically-thin plasma model.  The X-ray burst results from a sudden release of energy due to either a thermonuclear instability or accretion onto a neutron star, and  lasts  for   10$^1$--10$^2$ seconds. Its spectrum is typically approximated by a blackbody model (Lewin et al., 1993).  The tidal disruption event is a phenomenon during which a star is disrupted by a tidal force of a nearby massive black hole, resulting in brightening for several months (Gezari, 2021). A supernova occurs due to the gravitational collapse of a massive star or gas accretion onto a white dwarf. The timescale of a supernova is several weeks  to months (Dallal $\&$ Azzam, 2021).  The X-ray afterglow is a phenomenon that persists for several days following a gamma-ray burst (Galli et al. 2008). \par

The timescales of NuSTAR J230059+5857.4 are consistent with those of X-ray bursts and stellar flares but none of the others,  suggesting that NuSTAR J230059+5857.4 is likely to be associated with an X-ray burst or a stellar flare.  The spectrum of NuSTAR J230059+5857.4 is not  characterized with a blackbody model as in  X-ray bursts but with  an optically-thin thermal plasma model.   In addition, the SED obtained from  optical and infrared observations of the likely counterpart source of NuSTAR J230059+5857.4 can be  reproduced by  a stellar radiation model (BT-NextGen model). For these reasons, we conclude that NuSTAR J230059+5857.4 is most likely to be a stellar flare.\par

\subsection{Type of the star}
\label{Type of the star}

In this section, we  discuss the type of the star  from which NuSTAR J230059+5857.4 originated. The observed stellar radii shown in Table \ref{tab:opIRfitres} are approximately 10--100 times smaller than giants  in RS CVn systems. This suggests that the star responsible for the observed flare is unlikely to be an RS CVn system. \par

As described in section \ref{Optical and infrared observations}, the flare source could be a single M-dwarf or an M-dwarf binary. On the other hand, {\it GAIA} classified this star as not a binary system but a single star based on astrometry, spectroscopy, and the absence of eclipses.  At first glance, this fact appears to disprove the possibility that the flare source is a binary system. However, {\it GAIA} cannot spatially resolve binary systems composed of stars with  equal fluxes  with a separation of  less than 700 mas (de Bruijne et al. 2015), which is translated into 200 AU  for the  \textit{GAIA}-measured distance of 281 pc. This fact may suggest that the binary system is in the face-on configuration and the binary separation is  too small (smaller than 200 AU) for {\it GAIA}  to resolve it spatially.  Given that a separation below 200 AU is common in  M-dwarf binary systems (e.g., Fischer $\&$ Marcy 1992), it is possible that the flare source is an M-dwarf binary system. \par

The analysis in section \ref{Optical and infrared observations} has indicated that a pre-main-sequence star is  not excluded from a potential candidate for the flare source. Pre-main-sequence stars are located on the Hayashi track in the Hertzsprung-Russell diagram and are often observed  inside molecular clouds, in which star formation often occurs. Pre-main-sequence stars are  included in the class of young stellar objects (YSOs). YSOs are classified into Classes 0--III  according to their age,  and pre-main-sequence stars generally correspond to Classes II--III. Among these classes, Class III YSOs are typically   10$^6$--10$^7$ years old. When pre-main-sequence stars  evolve from Class II YSOs to Class III, their surrounding accretion disks become thin,  and many of them become weak-lined T Tauri stars (WTTS), which contains old Class II YSOs and Class III YSOs.\par

We examined the potential presence of molecular clouds  surrounding the star using a column density to obtain supporting evidences for whether the flare source is a pre-main-sequence star, which is often found within a molecular cloud while an M-dwarf is usually found outside of a molecular cloud. The column density derived from the {\it Chandra} data was $5.9^{+3.1}_{-2.4}\times10^{20}~{\rm cm}^{-2}$ (Table~6). This  value is markedly smaller  than  a typical value of  $\sim10^{22}~{\rm cm}^{-2}$ for a source inside a molecular cloud. It suggests that the flare source cannot be unequivocally identified as the pre-main-sequence star. However, Class III YSOs are sometimes observed outside  molecular clouds. Possible  causes for  such a system include gravitational scattering by nearby stars,  transfer of high velocities from small cloudlets  in molecular clouds, and starbursts occurring near the end of the molecular cloud's lifetime (Feigelson $\&$ Montmerle 1999). Consequently, it is possible that  the flare-source star is classified as the Class III YSO, around which the molecular cloud has cleared. \par

\subsection{Scale of the flare}
\label{Scale of the flare}

\begin{figure*}[htb]
  \begin{minipage}{0.5\hsize}
  \includegraphics[keepaspectratio,scale=0.24, angle=0]{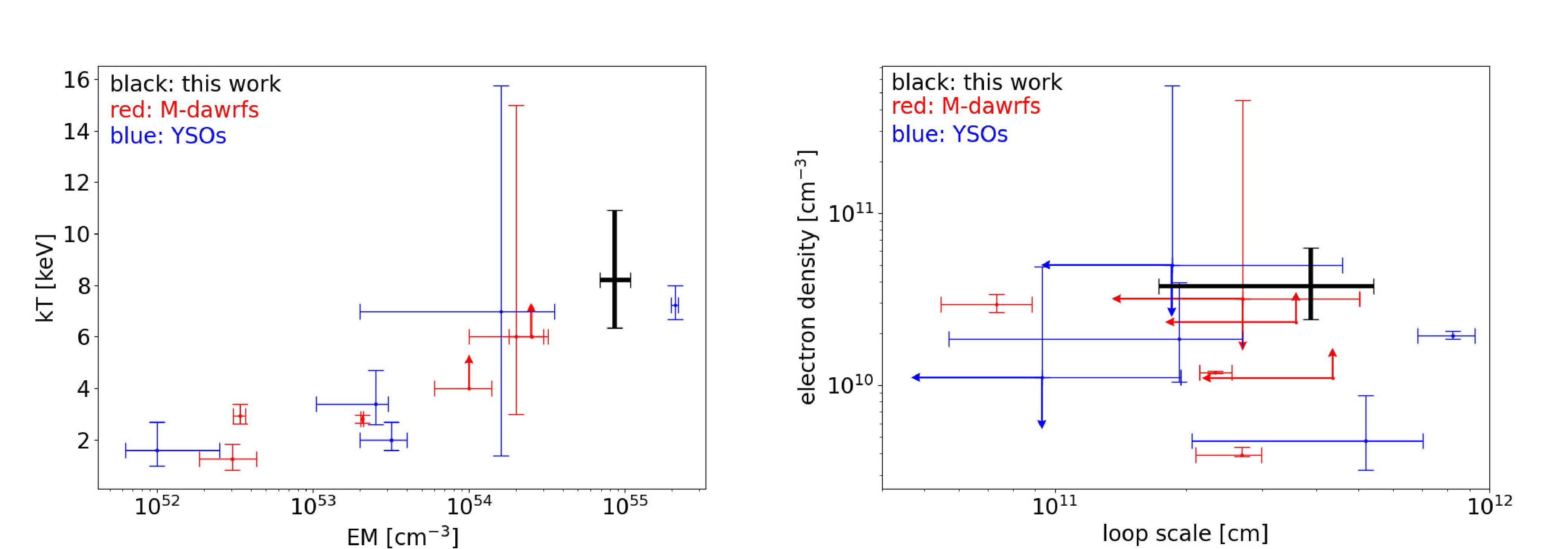}
 \end{minipage}
  \caption{Left panel: Comparison of the emission measures and temperatures  obtained in this work and previous  studies (Tsuboi et al. 1998, 2016; Favata et al. 2000a, 2000b; Imanishi et al. 2001; Wargelin et al. 2008). Right panel: Same as Left panel but of  the loop  sizes and electron densities. Crosses in black, red, and blue  show the data from this work, previous  studies on M-dwarfs and pre-main-sequence stars, respectively. Error bars represent  the 90$\%$ confidence interval.}
  
 \label{parconv}
\end{figure*}

  Figure \ref{parconv}  compares the emission measure, temperature, loop  size, and electron density of NuSTAR J230059+5857.4 with those of the M-dwarf and pre-main-sequence star flares discussed in previous studies. Most of the known pre-main-sequence stars fall into  Class III YSOs, but here we also include, for comparison purposes,  the Class II YSO V773 Tau.  V773 Tau is labeled as a WTTS, which has  been a subject of detailed analysis in many  studies, and thus  is a valuable  sample for our comparison. The \textit{NuSTAR} parameters  for the comparison are obtained through the analysis of the time-averaged spectra (Table \ref{tab:integspecfit}). For NuSTAR J230059+5857.4 and the previous work, the loop  sizes $L$ and electron densities $n$  are estimated  from  the emission measures $EM$, temperatures $T$, and  pre-flare electron density $n_{0}$  according to the theoretical flare model described in Shibata $\&$ Yokoyama (2002): 

\begin{equation}
\label{eq:L}
L = 5\times10^{11}\left(\frac{T}{10^{8}~{\rm K}}\right)^{-\frac{8}{5}}  \left(\frac{EM}{10^{55}~{\rm cm}^{-3}}\right)^{\frac{3}{5}}  \left(\frac{n_{0}}{10^{9}~{\rm cm^{-3}}}\right)^{-\frac{2}{5}}  ~\lbrack {\rm cm} \rbrack
\end{equation} 
and

\begin{equation}
\label{eq:n}
n = 10^{10}\left(\frac{T}{10^{8}~{\rm K}}\right)^{\frac{12}{5}}  \left(\frac{EM}{10^{55}~{\rm cm}^{-3}}\right)^{-\frac{2}{5}}  \left(\frac{n_{0}}{10^{9}~{\rm cm^{-3}}}\right)^{\frac{3}{5}}  ~\lbrack {\rm cm}^{-3} \rbrack
\end{equation} 

The units in the equations ((\ref{eq:L}), (\ref{eq:n}), and (\ref{eq:Eth}) mentioned later) are the typical values of NuSTAR J230059+5857.4. For the $EM$ and $T$, we used the values measured in each study. As the pre-flare electron densities  are difficult to estimate in some of the  observations, we compute the loop sizes and electron densities of the flares described in our work and previous work under an assumption of a fixed electron density of 10$^{9}$ cm$^{-3}$, a typical value for solar active regions. The loop length and electron density of V773 Tau were calculated as $8\times10^{11}$ cm and $2\times10^{10}$ cm$^{-3}$, respectively.\par

 As a result, we find that the temperature and emission measure of NuSTAR J230059+5857.4 are  higher  than those  for most of the flares from M-dwarf and pre-main-sequence stars reported in previous studies, with the emission measure being particularly pronounced (Figure~\ref{parconv}).  Although it is possible that such a hot and bright flare  may have indeed occurred, this result  might be attributed to sympathetic and multiple flares, which have been observed on the Sun (e.g., Moon et al. 2002; Titov et al. 2012). When the multiple flares occur synchronously on the star, the observed emission measure is the sum of multiple flares, and  the actual emission measure per flare is smaller than what is observed. Thus, $NuSTAR$ may have observed  multiple flares. \par

We find no significant differences in  flare loop  sizes and electron densities between the results of previous  studies and this  work.  We notice that the estimated loop  size in this work  of 4$\times~10^{11}$ cm is several times larger than the stellar radius.  Similarly, the estimated loop sizes in many of the   previous  studies  are several times larger than the stellar radius.  With solar flares, by contrast,  flare loops exceeding the size of the star have never been  observed.  Considering this discrepancy, we conjecture that   systematic overestimation  may be responsible for the observed apparently  large flare loops although it is still possible that very large loops are actually formed on many stars. In our analysis, we assumed  the pre-flare electron density  to be equal to that of the Sun. If it is  an underestimate, the loop  size is overestimated  because the loop  size is inversely proportional to the pre-flare electron density to the power of 2/5. \par

A correct interpretation of the measured $EM$ will be key to studying the flare property. If multiple flares occurred simultaneously during the observations, the observed emission measure will include the contributions from those multiple flares. If this is the case, a direct application of the flare scaling relations (equations (\ref{eq:L}) and (\ref{eq:n})) will be invalid as the relations are for a single flare. If we substitute $EM$ for multiple flares to the scaling relations, we will overestimate the flare loop size (see equation (\ref{eq:L})). Using the loop size, we estimated the total thermal energy, $E_{th}$, of the flare event. When the flare volume is estimated as $L^{3}$, the total thermal energy can be estimated as $3nk_{B}TL^{3}$, where $k_{B}$ is the Boltzmann constant. We therefore get

\begin{equation}
\label{eq:Eth}
E_{th} \sim 3\times10^{37}\left(\frac{T}{10^{8}~{\rm K}}\right)^{-\frac{7}{5}}  \left(\frac{EM}{10^{55}~{\rm cm}^{-3}}\right)^{\frac{7}{5}}  \left(\frac{n_{0}}{10^{9}~{\rm cm^{-3}}}\right)^{-\frac{3}{5}} \lbrack {\rm erg} \rbrack.
\end{equation}

Using the observed $EM$ and $T$, we calculate equation \ref{eq:Eth}, which results in E$_{th} \sim 10^{37}$ erg.  The estimated thermal energy is about 100 times larger than the measured total radiative energy in the X-ray ($\sim$ 10$^{35}$ erg. See Table \ref{tab:integspecfit}). Typically, the ratios between bolometric flare energy and X-ray radiative energy ranged approximately from 0.1 to 100 (Namekata et al. 2024). The obtained ratio is not unrealistic and represents one of the highest values among the ratios observed so far. If multiple flares occurred, the $EM$ for each individual flare would be several times smaller, leading to this ratio being reduced to the order of several tens. \par 

\subsection{Time variation of temperatures}
\begin{figure*}[htb]
  \begin{minipage}{0.5\hsize}
  \includegraphics[keepaspectratio,scale=0.25, angle=0]{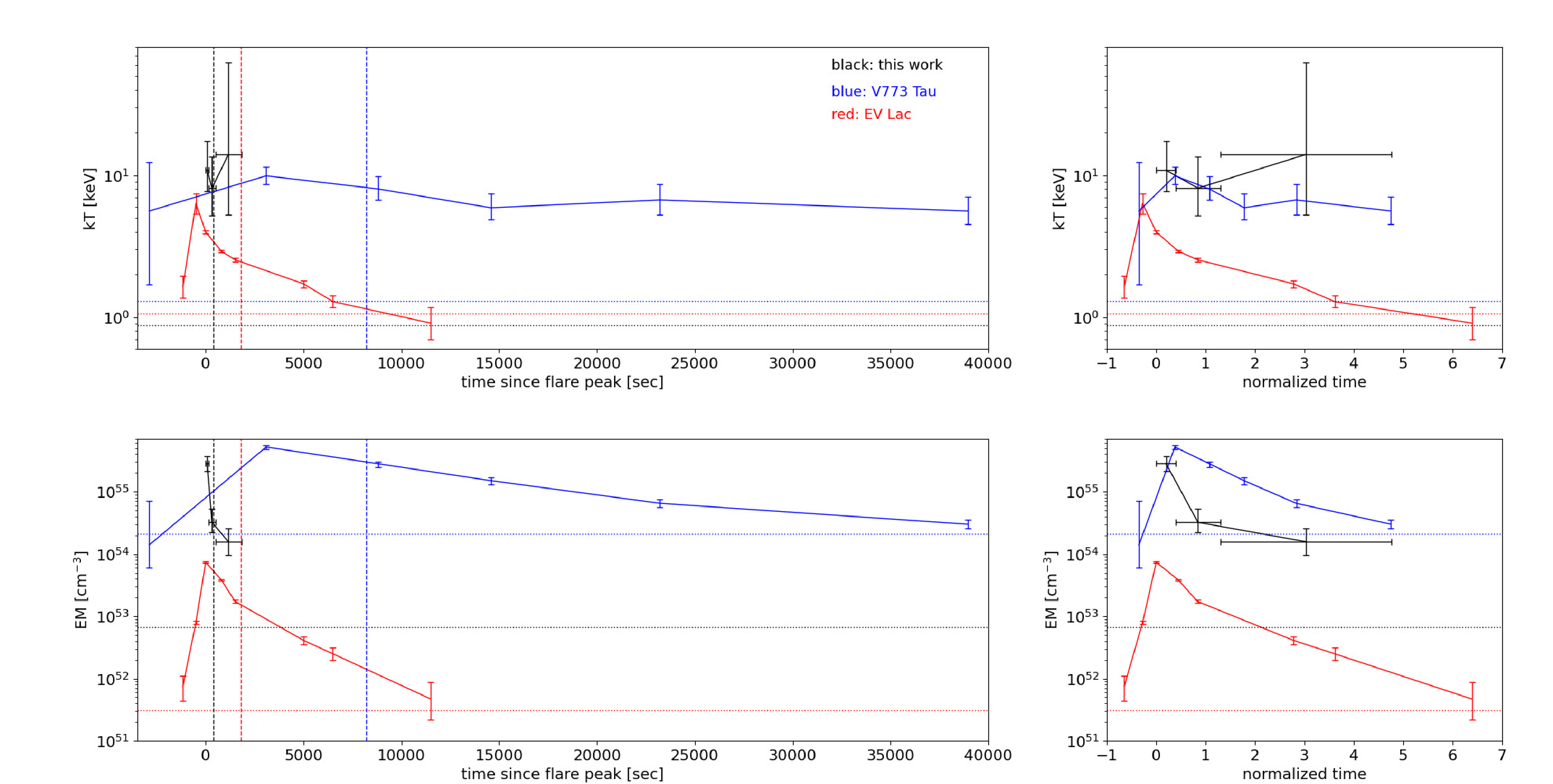}
 \end{minipage}
  \caption{Variation of the temperatures (upper two panels) and emission measures (bottom two panels) during the flares. In the left two panels, the horizontal axis represents the elapsed time in seconds from the flare peak, while one in the right two panels represents the elapsed time normalized by the e-folding time of each flare. Crosses and solid lines in black, red, and blue  show the data from this study, the M-dwarf EV Lac (Favata et al. 2000a), and the pre-main-sequence star V773 Tau (Tsuboi et al. 1998), respectively. The horizontal dotted lines represent the parameters in the quiescent phases for the respective sources. The vertical dashed lines represent the e-folding time of each flare. The vertical and horizontal error bars represent a 90$\%$ confidence interval and the time interval, respectively.}
 \label{emktval}
\end{figure*}

We discuss variations in temperature from the peak to the decay phase to  investigate the phenomena following the initial magnetic reconnection. Figure \ref{emktval}  shows the temperature and emission measure variations during the three flares: NuSTAR J230059+5857.4,   EV Lac flare (Favata et al. 2000a), and V773 Tau flare (Tsuboi et al. 1998). For NuSTAR J230059+5857.4, we use the   {\it NuSTAR}  parameters in phases (1)-(3) (Table \ref{tab:coplres}). As the e-folding time, we use the fitting result with a single exponential function for all the three sources because that was the only available results in Favata et al. (2000a) and Tsuboi et al. (1998). \par

As a result, we find  that   the temperature of NuSTAR J230059+5857.4 remains higher than that  in the quiescent phase for as long as or longer duration  than those of EV Lac and V773 Tau,  while the luminosity decreases following the initial magnetic reconnection. Favata et al. (2000a) and Tsuboi et al. (1998) suggested the possibility of additional heating occurring after the initial magnetic reconnection.  Our result indicates that the additional heating may also be present during the flare studied in this work. One possible explanation for the additional heating is that the  observed flare was sympathetic flares (multiple flares occurring sympathetically and consecutively). On the Sun,  sympathetic flares  have been observed to last for about 10 hours (Titov et al. 2012).  This fact suggests that $Swift$ observed  sympathetic flares occurring later. Future high time-cadence and high sensitivity monitoring observations will be able to test the hypothesis that sympathetic flares occurs on this star. Furthermore, it would help us determine whether the observed X-ray emission from multiple flares or a single flare. \par

\section{Summary}
\label{Summary}
(1) On 2013 April 25,  {\it NuSTAR}  detected  a transient source, NuSTAR J230059+5857.4, at (RA, Dec)$_{\rm J2000.0}$ = (\timeform{23h00m59.88s}$\pm$0$^{\rm s}$.65, \timeform{+58D57'25.86''}$\pm$6${\arcsec}\!.$14). The light curve was represented by  a  two-component decay with decay times of $\sim$70 and $\sim$1000 seconds. The time-averaged temperature was $\sim$8 keV, while it is peaked at $\sim$10 keV in the time-resolved spectral analysis\par

(2) {\it Chandra}, {\it XMM-Newton}, {\it Swift}, and {\it Suzaku} observed the  region where NuSTAR J230059+5857.4 emerged. The counterpart point source was detected in observations made by {\it Chandra}, {\it XMM-Newton}, and {\it Swift},  whereas it remained undetected in the {\it Suzaku} observation. Notably, {\it Swift} observed the decay phase immediately following the {\it NuSTAR} observation, while all the other observations were conducted during  a quiescent phase.\par

(3) In the optical and infrared bands, {\it GAIA}, {\it SDSS}, {\it 2MASS}, and {\it WISE} observed the object where NuSTAR J230059+5857.4 occurred. Based on the {\it GAIA} observation results, the distance to the object was determined to be 281 pc. From {\it SDSS}, {\it 2MASS}, and {\it WISE} data,  we found that the SED was  well reproduced by the BT-NextGen model with  a temperature  between 3400  and 4000 K and  a radius  between 0.73  and 1.04 R$_{\odot}$.\par

(4) We found that NuSTAR J230059+5857.4 was  likely to be a stellar flare  that originated from  a pre-main-sequence star, an M-dwarf system, or a single M-dwarf.  If the flare source is a pre-main-sequence star, it is  likely to be a Class III YSO on the basis of the {\it Chandra} X-ray result.   If the flare source  is an M-dwarf binary system, the separation  should be  smaller than 200 AU  on the basis of the {\it GAIA} data. \par

(5) The temperature and emission measure of the flare were larger than those reported in previous studies, with the emission measure being particularly significant. The estimated loop  size and electron density of the flare are consistent with the parameter distribution for past observed stellar flares.  However, given that the estimated loop  size is several times larger than the stellar radius, the loop size might be overestimated.  Indeed, either or both of  an underestimation of the pre-flare electron density and  an overestimation of the emission measure would lead to an overestimation of the loop size. The latter is consistent with the observed large emission measure, which may be caused by  observations of  multiple flares that occurred sympathetically. \par

(6) The temperature of the flare remained higher than that during the quiescent phase for a comparable to or longer duration than  those of previous  studies. This fact suggests the presence of additional heating following the initial magnetic reconnection, which  might be attributed to the continued occurrence of  sympathetic flares. \par

\begin{ack}
We thank M. Sakano for the useful discussions. This work was supported by JST SPRING, Grant Number JPMJSP2138, and JSPS KAKENHI Grant numbers 20H00175, 23H00128.
\end{ack}

\newpage


\end{document}